\documentclass[prl,aps,twocolumn,superscriptaddress,showpacs,tightenlines,floatfix,amsmath,amssymb]{revtex4}

\usepackage{bm}

\usepackage{amsmath}
\usepackage{graphicx}
\usepackage{amssymb}
\usepackage{mathrsfs}
\usepackage{bbm}

\newcommand{\be}{\begin{eqnarray}}
\newcommand{\ee}{\end{eqnarray}}

\newcommand{\me}{{\mathbbmss e}}

\newcommand{\cG}{{\cal G}}

\newcommand{\cL}{{\cal L}}

\newcommand{\vect}[2]{\left(\begin{array}{c} #1 \\[1mm] #2 \end{array}\right)}

\begin{document}

\title{Embedding Brans-Dicke gravity into electroweak theory}

\vskip .0cm
\author{M. N. Chernodub}
\email{Maxim.Chernodub@itep.ru}
\affiliation{Institute of Theoretical and Experimental Physics,
B. Cheremushkinskaya 25, Moscow 117218, Russia}
\author{Antti J. Niemi}
\email{Antti.Niemi@teorfys.uu.se}
\homepage{http://www.teorfys.uu.se/people/antti}
\affiliation{
Laboratoire de Mathematiques et Physique Theorique
CNRS UMR 6083, F\'ed\'eration Denis Poisson, Universit\'e de Tours,
Parc de Grandmont, F37200, Tours, France}
\affiliation{Department of Theoretical Physics,
Uppsala University,
P.O. Box 803, S-75108, Uppsala, Sweden}
\affiliation{ Chern Institute of Mathematics,
Tianjin 300071, P.R. China }

\preprint{UUITP-15/07}
\preprint{ITEP-LAT/2007-01}

\begin{abstract}
We argue that a version of the four dimensional Brans-Dicke theory
can be embedded in the standard flat spacetime electroweak theory.
The embedding involves a change of variables that separates the
isospin from the hypercharge in the electroweak theory.
\end{abstract}


\date{\today}

\maketitle



%
%
\vskip 1.cm
The construction of a consistent four dimensional
quantum theory of gravity remains a challenge. Superstring theory
with its pledge to unify all known interactions is
the most attractive candidate for resolving this conundrum \cite{polbook}.
Some colleagues have also argued that conformal Weyl gravity
is a renormalizable albeit not apparently unitary
four dimensional quantum theory of gravity \cite{weyl}.
Finally, there are indications that $\mathcal N=8$ supergravity
theory might be ultraviolet finite, for reasons
that resemble those ensuring the finiteness of
the $\mathcal N=4$ supersymmetric Yang-Mills theory \cite{bern}.
The spectrum of the latter appears to
coincide with that of the $AdS_5\times \mathbb S^5$
solution of ten dimensional $IIB$ supergravity theory \cite{mald}.
This duality has led \cite{pol} to propose that
even within the strong and the electroweak components of
the Standard Model there is an embedded quantum theory of
gravity that remains to be discovered. (See also~\cite{volovik}.)

\vskip 0.3cm
In the present paper we inspect how a gravity theory
could be embedded in the bosonic sector of the (Euclidean
signature) Weinberg-Salam Lagrangian \cite{abers},
\begin{equation}
\cL_{WS}^{}
= \frac{1}{4} \vec F^{\hskip 0.6mm 2}_{\mu\nu} + \frac{1}{4} B^2_{\mu\nu}
+ | D_\mu { \phi}|^2
+  \lambda ({ \phi}^\dagger { \phi})^2 +
\mu^2  \phi^\dagger \phi
\label{ew:lag}
\end{equation}
All our notations are {
exactly} as in \cite{abers}.
We decompose the Higgs field $\phi$
and the $SU_L(2)$ gauge field $A^a_\mu$ as follows,
\begin{eqnarray}
\phi  & = & \left(  \sigma \, e^{i\alpha} \right ) \cdot
\mathcal S\,,
\quad
\mathcal S =
\frac{e^{i\gamma}}{ \sqrt{2(1-n_3)}} \vect{n_1-in_2}{1-n_3}
\label{eq:decom} \\
\widehat A_\mu & = &
\left\{ \left( \mathcal W_\mu^3 + \frac{2i}{g} \,
\vec m^{+} \! \cdot \partial_\mu \vec m^{-}
\right) {\hat n} \ - \ \frac{i}{2g}
\hskip 0.6mm \left[ \, \partial_\mu  {\hat n} \, , \,  {\hat n} \, \right]
\right\}\nonumber\\
& & + \, \left\{ \mathcal W_\mu^+ \cdot {\hat m}^+
\ + \ \mathcal W_\mu^- \cdot {\hat m}^-\right\} \ \equiv \
 \hat {\mathcal A}_\mu \, + \, \hat{\mathcal X}_\mu
\label{cB}
\end{eqnarray}
Here $\hat n$ is the isospin projection operator
\[
\hat n \ = \ \vec n \cdot \vec \tau  \ = \ - \,
\frac{ \phi^\dagger \vec \tau \phi}{\phi^\dagger \phi} \cdot \vec
\tau
\]
and $\vec m^\pm = \vec m^1 \pm i \vec m^2$ so that $(\vec m^1 ,
\vec m^2, \vec n)$ is a right-handed orthonormal triplet.
If the phases $\alpha$ and $\gamma$ are
combined, (\ref{eq:decom}), (\ref{cB}) involves
sixteen independent fields as a complete decomposition should. But
for future reference
we prefer to keep $\alpha$ and $\gamma$ separate.
Notice that $\vec m^\pm$ is defined up to a phase that we
may identify with $\gamma$.
We have selected (\ref{eq:decom}) and (\ref{cB})
so that in the unitary gauge where $\hat n$ becomes
equal to the diagonal Pauli matrix
$\tau^3$, the diagonal
component of the gauge field coincides with $\mathcal W^3_\mu$
and the off-diagonal components coincide with
the $\mathcal W_\mu^\pm$. In the low-temperature
Higgs phase $\mathcal W^3_\mu $ then
combines with the $U_Y(1)$ hypergauge field $B_\mu$ into the massive
neutral $Z$-boson and the massless photon,
and the off-diagonal gauge fields $\mathcal W^\pm_\mu =
\frac{1}{\sqrt{2}} \Bigl(\mathcal W^1_\mu \mp i\mathcal W^2_\mu\Bigr) $
become the massive charged $W$-bosons \cite{abers}.

In the present paper we shall argue that a gravity theory
emerges from (\ref{ew:lag}) in terms of spin-charge decomposed variables.
For this we start by noting
that the Higgs field is a Lorentz scalar but with a nontrivial isospin and a
nontrivial hypercharge. Thus we decompose it accordingly, and
the result is displayed in (\ref{eq:decom}). Since the physical
$Z$-boson and photon
are both charge neutral,  there is no room for any
spin-charge decomposition in $\mathcal W^3_\mu$ and $B_\mu$.
But the $W$-bosons are charged and
they also have a nontrivial Lorentz spin, and
following \cite{faddeco} we separate their spin from their
charge by decomposing
\begin{equation}
\mathcal W^+_\mu = \psi^{}_1 {\me}^{}_\mu +
\psi^{}_2 \bar{\me}^{}_\mu
\label{eq:xi:psi}
\end{equation}
The $\psi^{}_{1,2}$ are two complex scalars,
they carry the charge of $\mathcal W^\pm_\mu$ \cite{faddeco}.
The complex four-vector ${\me}^{}_\mu$ carries spin,
it is normalized according to
\begin{equation}
{\me}^{}_\mu {\me}^{}_\mu = 0 \ \ \ \ \ \& \ \ \ \ \
{\me}^{}_\mu  \bar{\me}^{}_\mu = 1
\label{eq:e:bare}
\end{equation}

The decomposition (\ref{eq:decom}) admits an internal
$U_\phi(1)$ gauge symmetry: If we
send $\alpha \to \alpha + \delta$ and $\gamma \to
\gamma - \delta$ the Higgs field $\phi$ remains intact. Similarly,
if we multiply $\psi_{1} $ and $\psi_{2}$
by a phase and $\me_\mu$ by the complex conjugate phase,
(\ref{cB}) and (\ref{eq:xi:psi}) do not change
under this internal $U_W(1)$ gauge transformation.
The gauge fields for these internal symmetries are composite
vector fields. For $U_\phi(1)$ we have $
\Lambda_\mu = - i \mathcal S^\dagger \partial_\mu \mathcal S$
and for $U_W(1)$ we have $ \mathcal C_\mu = i \bar\me \cdot \partial_\mu \me$.

We now define a number of auxiliary quantities.
We start by introducing the three component unit vector
\[
\vec t  \ = \ \frac{1}{\rho^2}\cdot
\left( \begin{matrix} \bar\psi^{}_1 & \bar\psi^{}_2 \end{matrix} \right)
\vec \sigma \left( \begin{matrix} \psi^{}_1 \\
\psi^{}_2 \end{matrix} \right)
\]
where $\rho^2 =|\psi_1|^2 + |\psi_2|^2$. With
$ g \mathcal Y_\mu = g \hskip 0.6mm \vec n \cdot \vec{\mathcal A}_\mu -
2\Lambda_\mu$ we define a $U_\phi(1) \times U_W(1) $
covariant derivative as follows,
\begin{eqnarray}
\mathrm D^{\mathcal C}_{\mu}\psi^{}_{1,2} & =
& (\partial_\mu + i g \mathcal Y_\mu \mp i \mathcal C_\mu) \psi^{}_{1,2}\,,
\quad
\nonumber\\
\mathrm D^{\mathcal C}_{\mu} \me^{}_\nu
& = &
(\partial_\mu + i \mathcal C_\mu) \me_\nu
\nonumber
\end{eqnarray}
and a $U_\phi(1) \times U_Y(1)$ covariant derivative
as follows,
\[
\mathfrak D_\mu = \partial_\mu - i \frac{g}{2} \mathcal
Y_\mu - i \frac{g'}{2} B_\mu
\]
We then introduce the gauge invariant supercurrents
\begin{eqnarray}
J^{\pm}_\mu & = & \frac{i}{2\rho^2}
\{ \left( \bar\psi^{}_1  \mathrm D^{\mathcal C}_{\mu}
\psi_1 - \psi^{}_1 \bar {\mathrm D}^{\mathcal C}_{\mu}
\bar\psi^{}_1\right) \pm
(1 \to 2) \} \nonumber\\
T_\mu \ & = & \ - \frac{i}{2\sigma^2} \{ \phi^* \mathfrak D_\mu \phi - \phi
\, \bar{\mathfrak D}_\mu \phi^* \}
\end{eqnarray}
Finally,
\[
P_{\mu\nu}  =   \frac{1}{2} ig \cdot \rho^2 t^{}_3
\cdot(\me^{}_\mu \bar{ \me}^{}_\nu -
\me^{}_\nu \bar{\me}^{}_\mu ) \equiv
g\rho^2 t_3{H}_{\mu\nu}\,.
\]
Following \cite{faddeco}
we interpret $\rho^2$ as the conformal
scale of a locally conformally flat metric tensor,
\begin{equation}
G_{\mu\nu} \ = \ \left(\frac{\rho}{\kappa}\right)^2 \delta_{\mu\nu}
\label{met}
\end{equation}
and {
from now on} all the Greek indices $\mu,\nu,\lambda, ...$
refer to the ensuing locally conformally flat spacetime.
Note that in a coordinate basis
the metric tensor is
dimensionless while $\rho$ has the dimensions of mass. Dimension
analysis then tells us to introduce
the {\it a priori} arbitrary mass parameter $\kappa$.
With the metric tensor we have the vierbein ${E^a}_\mu$
that relates a coordinate basis ($\mu$) to a local orthogonal
frame ($a$), the Christoffel symbol
$\Gamma^{\lambda}_{\mu\nu}$, the spin connection
$\omega^{\, \, \lambda}_{\mu \,\, \nu}$ and all other geometric
quantities that are defined in the usual, standard fashion.
The covariant derivative
of the zweibein field ${\me}^{}_\mu $ is \cite{faddeco}
\[
\nabla_\mu {\me}^{}_\nu  + \omega^{\, \, \lambda}_{\mu \,\,
\nu}{\me}^{}_\lambda = \partial_\mu{\me}^{}_\nu  -
\Gamma^{\lambda}_{\mu\nu} {\me}^{}_\lambda
+ \omega^{\, \, \lambda}_{\mu \,\, \nu} {\me}^{}_\lambda
=
\rho \cdot \partial_\mu ( \frac{ {\me}^{}_\nu }{\rho})
\]
The covariantized $U_W(1)$ gauge field is
\[
{\mathcal C}_\mu  =
i {\bar{\me}}^{\sigma} \nabla_\mu {\me}_\sigma
+ i \bar{\me}^{\lambda} \omega^{\,\, \sigma}_{\mu \, \,
\lambda}  {\me}^{}_\sigma
\]
and we also introduce the following
twisted covariant derivative operator
\[
{\mathcal D}_{\mu \,\, \lambda}^{\,\, \nu}
= \delta_{ \,\,\, \lambda}^{\nu} \nabla^{\mathcal C}_\mu +
\omega^{\, \, \nu}_{\mu \,\, \lambda}
=   \delta_{ \,\,\, \lambda}^{\nu} (\nabla_\mu + i \mathcal C_\mu)
+ \omega^{\, \, \nu}_{\mu \,\, \lambda}
\]
Finally, in $4D$ the Ricci scalar for our
metric tensor is
\[
R = - 6 \left( \frac{\kappa}{\rho}\right)^2 \left\{
\, \frac{1}{\rho^2} (\partial_\mu \rho)^2 \,
+ \, \partial_\mu ( \frac{1}{\rho} \, \partial_\mu \rho )
\, \right\}
\]

In order to relate (\ref{ew:lag}) to a gravity theory, we
first employ the present geometrical structure
to convert it into a generally covariant
form. The result is a sum of two
terms $\cL^{}_{WS} = \cL^{(1)}_{WS} +
\cL^{(2)}_{WS} $ that we now inspect separately.
We start with $\cL^{(1)}_{WS}$. It admits two contributions,
the first one is
\begin{eqnarray}
& &
\hskip -7mm
\cL_{WS}^{(11)} =
\frac{1}{4} \sqrt{G} G^{\mu\nu}
G^{\rho\sigma} \mathfrak G_{\mu\rho}\mathfrak G_{\nu\sigma}
\label{fin:1} \\
& &
\hskip -7mm
+ \frac{1}{4} \sqrt{G} G^{\mu\nu}
G^{\rho\sigma} \mathcal T_{\mu\rho}\mathcal T_{\nu\sigma}
+ \varsigma^2 \sqrt{G} G^{\mu\nu} T_\mu T_\nu
\label{fin:2}
\\
& &
\hskip -7mm
+ \sqrt{G} G^{\mu\nu} \partial_\mu \varsigma \partial_\nu
\varsigma + \frac{ \varsigma^2 }{6} \cdot R \, \sqrt{G}
+ \sqrt{G} \{ \lambda \hskip 0.3mm \varsigma^4 + \mathit r \varsigma^2
\, \}
\label{fin:3}
\end{eqnarray}
We have defined
\begin{equation}
\mathfrak G_{\mu\nu} =
\mathcal G_{\mu\nu} (\mathcal C) - (\partial_\mu J^+_\nu - \partial_\nu
J^+_\mu) - 2g^2 {\kappa}^2 t_3 { H}_{\mu\nu}
\label{frakG}
\end{equation}
where $\mathcal G_{\mu\nu} (\mathcal C)$ is the 't Hooft tensor \cite{thoof}
\begin{equation}
\mathcal G_{\mu\nu}(\mathcal C)
= \partial_\mu [ t_3 {\mathcal C}_\nu ]
- \partial_\nu [ t_3 {\mathcal C}_\mu  ] - \frac{1}{2}
\vec t \cdot \partial_\mu \vec t \times \partial_\mu \vec t
\label{tho}
\end{equation}
and
\[
\mathcal T_{\mu\nu} =
\frac{1}{g'} \left[ \cG_{\mu\nu}(\mathcal C) - (\partial_\mu J^+_\nu
- \partial_\nu J^+_\mu) -  2 (\partial_\mu T_\nu - \partial_\nu T_\mu)
\right]
\]
and $ \varsigma = G^{-1/8} \hskip 0.5mm \sigma$ and
$\mathit r = G^{-1/4} \mu^2$.
Note in particular that (\ref{fin:1})-(\ref{fin:3}) have no
explicit $\kappa$ dependence except for the last term in
$\mathfrak G_{\mu\nu}$.

The second contribution to  $\cL^{(1)}_{WS}$ is
\begin{eqnarray}
& &
\hskip -7mm
\cL_{WS}^{(12)} = \kappa^2 \cdot \, { \left \{ { \,
\frac{1}{2}\, \sqrt{G} \, G^{\mu\nu} \left[ \,
J^+_\mu J^+_\nu
+ \frac{1}{4}
\nabla^{\mathcal C}_\mu \vec t \cdot \nabla^{\mathcal C}_\nu
\vec t \right. } \right. }
\label{fin:4}
\\
& &
\hskip -7mm
{ \left.
+ ({\bar {\mathcal D}}_{\mu \,\, \lambda}^{\,\,
\sigma} \bar{\me}^{}_\sigma)
({\mathcal D}_{\nu \,\, \eta}^{\,\, \tau} {\me}^{}_\tau)
+ \frac{1}{2} t_-
({{\mathcal D}}_{\mu \,\, \lambda}^{\,\,
\sigma} {\me}^{}_\sigma) ({{\mathcal D}}_{\nu \,\,
\eta}^{\,\, \tau} {\me}^{}_\tau) + c.c. \right] }
\label{fin:5}
\\
& &
\hskip -7mm
\left. + \frac{1 }{12}\cdot R \, \sqrt{G}
+ \frac{1}{4} g^2 ( \varsigma^2  -
\frac{3}{8}\hskip 0.3mm t^2_3 {\kappa}^2 \hskip 0.3mm
\} \cdot \sqrt{G} \, \right \}
\label{fin:6}
\end{eqnarray}
Notice that the {entire}
(\ref{fin:4})-(\ref{fin:6}) is proportional to $\kappa^2$.

Before we proceed to the final term
\begin{equation}
\cL^{(2)}_{WS} \, = \, -
\frac{1}{2}  (D^{ab}_\mu [\mathcal A] \mathcal X_{\mu b})^2
\label{sb}
\end{equation}
we first point out some salient features in the structure
of (\ref{fin:1})-(\ref{fin:3}),
(\ref{fin:4})-(\ref{fin:6}):

We start by observing that $ \cL^{(1)}_{WS}$
involves only $SU_L(2)\times U_Y(1)$
gauge independent variables. There are sixteen independent fields,
in addition to the $U_W(1)$ phase.

Paramount to our geometric interpretation of $ \cL^{(1)}_{WS}$
is that the density $\rho$
is a nonvanishing quantity, $\langle \, \rho \, \rangle =
\Delta \not= 0$. {\it A priori} it could be natural to
identify $\Delta \equiv \kappa$
but we keep them separate. Arguments have been given \cite{zakh},
\cite{zakh2} that in a $SU(2)$
Yang-Mills theory $\Delta$ is nonvanishing. Assuming that this persists
in the Weinberg-Salam model, the Lagrangian (\ref{fin:1})-(\ref{fin:3}),
(\ref{fin:4})-(\ref{fin:6}) is defined in a locally conformally
flat spacetime which is {different} from
the flat $\mathbb R^4$ of perturbation theory. This emergence of a novel
spacetime is in line with the {\it no-go} theorem \cite{wewi},
\cite{com} that forbids an embedded theory of gravity from residing
in the {same} spacetime with the underlying nongravity theory;
see also \cite{pol}.

The contribution (\ref{fin:6}) has the standard Einstein-Hilbert
form with a cosmological ``constant'' term.
Similarly the two first terms in  (\ref{fin:3}) constitute a Brans-Dicke
Lagrangian: With $\psi = \varsigma^2$ we arrive at the
standard Brans-Dicke form \cite{brans} with the conformally
invariant parameter value $\omega = -\frac{3}{2}$.

The contribution (\ref{fin:1}) together with the first term
in (\ref{fin:4}) describe the embedded dynamics
of the supercurrent $J^+_\mu$ with mass $\kappa$.
The kinetic term of $J^+_\mu$ is also
embedded in the first term of (\ref{fin:2}).
Similarly, the two terms in (\ref{fin:2}) describe the embedded
dynamics of the supercurrent $T_\mu$. It
becomes massive whenever we are in a Higgs phase where
$\varsigma$ acquires a nontrivial expectation value.

It is notable that when $\kappa \not= 0$
the vector $J^+_\mu$ acquires a mass
even in the absence of the conventional Higgs effect.

Both $ \mathfrak G_{\mu\nu}$ and  $\mathcal T_{\mu\nu}$ contain
the 't~Hooft tensor (\ref{tho}). Together with the kinetic term
in (\ref{fin:4}) for the vector field $\vec t$ this gives us an
embedded, unitary gauge version of the spontaneously
broken $SO(3)$ Georgi-Glashow model. The unbroken symmetry
group is the {\it compact} $U_W(1)$ that has the capacity of
supporting embedded
magnetic monopoles.

The second term in (\ref{tho}) in combination with the
$\vec t$ contribution in (\ref{fin:4})
describes the embedding of the Faddeev model \cite{fadori}.
Consequently we expect embedded knotted solitons \cite{nature} to be
present. Furthermore, in a Lorentz invariant ground state we
must have $t_3 = \pm 1$  \cite{faddeco} and this prevents $\vec t$
from supporting any massless modes.

The contribution (\ref{fin:5}) and the
last term in $\mathfrak G_{\mu\nu}$ defines
a (gauged) Grassmannian nonlinear sigma-model. Its properties
are detailed in \cite{faddeco}, \cite{marsh}.
Together the unit vector $\vec t$ and the complex vector $\me_\mu$
describe a six dimensional internal space with
the structure of $\mathbb S^2
\times \mathbb S^2 \times \mathbb S^2$.

The last term in (\ref{fin:3}) and the
second term in (\ref{fin:6}) combine into a
cosmological ``constant'' contribution. The original constant parameter
$\mu^2$ has become a spacetime dependent variable $\mathit r$.
We can interpret it as a background scalar curvature and
we can combine it with the middle term in (\ref{fin:3}). In addition,
the Brans-Dicke-Higgs field $\varsigma$ has a mass term
which is proportional to $\kappa$.
This mass together with the scalar
curvature $R$ in (\ref{fin:3}) influence how symmetry becomes broken
by the Higgs potential. In particular, there can be regions in
the spacetime where the symmetry is broken while in other
spacetime regions symmetry remains unbroken \cite{fw}.

In the vicinity of the (Lorentz invariant \cite{faddeco})
$t_3 = \pm 1$ ground state and
when the field variables are slowly varying, we may delete all
derivative contributions to the Lagrangian (\ref{fin:1})-(\ref{fin:3}),
(\ref{fin:4})-(\ref{fin:6}). We also assume
that $\mathit r$ describes the entire ground state
scalar curvature $\langle R \rangle$. When we minimize the ensuing
potential for $\varsigma$  in (\ref{fin:6}) and account
for the $H_{\mu\nu}$ in the first term
of (\ref{fin:1}), we conclude that
the (classical level) cosmological constant becomes
vanishingly small when the background scalar curvature
$\mathit r \sim \langle R \rangle$ is
\begin{equation}
\mathit r \cdot \Delta
\ = \
\mu^2 \approx  - \left( \sqrt{ \frac{\lambda}{2} } g + \frac{g^2}{4}
\right) \Delta
\label{cosmo}
\end{equation}
This gives
\begin{equation}
\langle \varsigma^2 \rangle
\approx \frac{1}{2\sqrt{2}} \frac{g}{\sqrt{\lambda}} \cdot \kappa^2
\label{exp}
\end{equation}
Suppose now that (\ref{cosmo}), (\ref{exp}) hold and
that we are near the BPS limit so that $\lambda$
is vanishingly small, and that
$\kappa$ is finite ({\it e.g.} of the order of the electroweak scale). The
cosmological constant then vanishes and the
effective Planck's mass in the second term of (\ref{fin:3})
can become very large. The vector fields $J_\mu$
and $\vec t$ both have a mass which is
of the order of the electroweak scale, but the vector field
$T_\mu$ becomes very massive.

Finally, in the London limit
where $\rho$ is constant, $ \cL^{(1)}_{WS}$
describes the interactions between
$J^\pm_\mu, \, T_\mu, \, \vec t$, $\me_\mu$ and
$\varsigma$ in a flat spacetime which is {different} from
the flat $\mathbb R^4$ where (\ref{ew:lag}) is defined.

We now proceed to the remaining contribution
(\ref{sb}). Notably it is independently $SU_L(2) \times U_Y(1)$ gauge
invariant. It describes the interactive dynamics
between the Grassmannian vector field ${\me}^{}_\mu$ and
the two complex currents
\[
\mathcal J^{(\pm)}_\mu \ = \ \frac{1}{4} \Gamma^\nu_{\nu\mu}
+ \frac{1}{2} \partial_\mu \ln (1\pm t_3)
- \frac{i}{t_3 \pm 1}
\cdot (\,
J^+_\mu \pm J^-_\mu \, )
\]
It also acquires a form which is generally covariant {\it w.r.t.} the
metric tensor (\ref{met}). In particular, in parallel with
(\ref{fin:4})-(\ref{fin:6})
the {entire} contribution (\ref{sb}) is proportional to $\kappa^2$.

Since $\mathcal J^{(\pm)}_\mu$ contains $\Gamma^\nu_{\nu\mu}$,
the presence of
(\ref{sb}) breaks the invariance
under four dimensional diffeomorphism group {\it Diff(4)}
into  {\it S\!Diff(4)}, its volume preserving subgroup.
But we have also observed that when $\kappa \not= 0$ both vector
fields $J_\mu^\pm$ and $T_\mu$ are massive in the Higgs phase.
As a consequence whenever $\kappa \not= 0$ the physical
photon field becomes subject to the Mei\ss ner effect and acquires a mass
in the Higgs phase. But since the photon mass (if there is any!) is
tiny \cite{pdg}, in order for us to reconcile with the observed
Physics we must take the limit $\kappa \to 0$. Since both
(\ref{fin:4})-(\ref{fin:6}) and (\ref{sb}) are proportional
to $\kappa^2$ this truncates the entire Lagrangian into
(\ref{fin:1})-(\ref{fin:3}).

We are now in the position to state the main proposal of
the present paper: {\it When the metric
tensor} $G_{\mu\nu}$ {\it in the Lagrangian} (\ref{fin:1})-(\ref{fin:3}) {\it
is taken to be arbitrary, this Lagrangian describes the gravity theory
that emerges from the electroweak Lagrangian} (\ref{ew:lag}).

Since the 't~Hooft tensor (\ref{tho})
is closed it can be written as the exterior
derivative of a (generally singular)
vector field, and this vector field can be combined
with $J^+_\mu$. Thus both $\vec t$ and $\me_\mu$ entirely disappear
from (\ref{fin:1})-(\ref{fin:3})
when $\kappa \to 0$. When the metric tensor is arbitrary but
{\it Diff(4)} symmetry remains broken into
volume preserving {\it SDiff(4)} as $\kappa \to 0$, the
Lagrangian (\ref{fin:1})-(\ref{fin:3}) with a {\it a priori}
arbitrary metric tensor engages sixteen {\it SDiff(4)} invariant
fields and this coincides exactly with the number
of $SU_L(2) \times U_Y(1)$ invariant fields
in (\ref{ew:lag}): The gravity Lagrangian
describes the interactive dynamics of the conformal
Brans-Dicke theory with a massless $J^+_\mu$ and with a $T_\mu$
that acquires a mass in the Higgs phase of the Brans-Dicke-Higgs
scalar field $\varsigma^2$.

The $\kappa\to 0$ limit is like a Wigner-In\"on\"u contraction: The
identification (\ref{met}) becomes singular but at the level of the Lagrangian
(\ref{fin:1})-(\ref{fin:3}), (\ref{fin:4})-(\ref{fin:6}), (\ref{sb})
with the {\it arbitrary} and in particular
$\kappa$-independent metric, the $\kappa \to 0$
limit is well defined.

Finally, we propose the following interpretation for the
appearance of a locally conformally flat metric tensor in the original
Lagrangian (\ref{fin:1})-(\ref{fin:3}): We view this Lagrangian
as the short distance limit of a higher derivative (one loop)
renormalizable gravity theory~\cite{weyl} with the additional term
\[
\Delta \cL \sim \frac{1}{4\gamma^2} \cdot W^2_{\mu\nu\rho\sigma}
\]
where $W_{\mu\nu\rho\sigma}$ is the Weyl tensor. Since the coupling
$\gamma$ is asymptotically free, the $\beta$-function
for $\gamma$ enforces the Weyl tensor to vanish at
the short distance limit \cite{weyl}. This
reduces the general metric tensor into its locally
conformally flat form in the short distance limit; see
also \cite{faddeco}.

In conclusion, we have constructed a change of variables that converts the
bosonic Weinberg-Salam Lagrangian into a variant of the Brans-Dicke Lagrangian
in a locally conformally flat spacetime. We have argued
that when the metric tensor becomes arbitrary
and we take the limit where the physical photon
becomes massless, this Brans-Dicke Lagrangian determines the
gravity theory which is embedded in the Weinberg-Salam Lagrangian.
We  expect that one can similarly relate a gravity theory to the
strong sector of the Standard Model. It would be interesting
to work out the details in particular since the enlarged
structure of the $SU(3)$ gauge group may directly engage
the remaining components of a full metric tensor.
We leave it as a puzzler to physically interpret
the possibility that within the Standard Model there may
be two distinct embedded gravity theories
with their own distinct spacetimes.

\vskip 0.5cm

\begin{acknowledgments}
This work has been supported by a STINT Institutional grant IG2004-2 025.
The work by M.N.Ch. is also supported by the
grants RFBR 05-02-16306a and RFBR\--DFG 06-02-04010.
The work by A.J.N is also supported
by a VR Grant 2006-3376 and by the Project Grant
$\rm ANR~ NT05-1_{}42856$. The authors
are most grateful to L. Faddeev for numerous discussions,
valuable comments and criticism. We also thank
S. Slizovskiy, F. Wil\-czek, M. Zabzine and K. Zarembo for discussions.
M.N.Ch. is thankful to the members of
Department of Theoretical Physics of Uppsala University,
Institute for Theoretical Physics of Kanazawa University,
and Laboratoire de Mathematiques et Physique Theorique of Tours University
for hospitality and stimulating environment. A.J.N. thanks the
Aspen Center for Physics for hospitality.
\end{acknowledgments}

\end{document}